# Multi-Target Maneuver Coordinations: Unlocking Coordination Opportunities in Connected Automated Driving

Rafael Molina-Masegosa[a], Sergei S. Avedisov[b], Miguel Sepulcre[a], Takayuki Shimizu[b], Javier Gozalvez[a], Onur Altintas[b]
[a] Networked Systems Lab, Universidad Miguel Hernandez de Elche, Elche, Spain, {rafael.molinam, msepulcre, j.gozalvez}@umh.es
[b] Toyota Motor North America R&D – InfoTech Labs, Mountain View, CA, USA, {sergei.avedisov, takayuki.shimizu onur.altintas}@toyota.com

***Abstract*— Maneuver coordination is a key enabler of connected and automated driving, allowing vehicles to negotiate and execute maneuvers that would otherwise be difficult, inefficient or unsafe. Existing approaches and use cases typically assume coordination with a single predefined target vehicle, which limits the number of coordination opportunities. This paper introduces a maneuver coordination approach based on multi-target selection, which allows a vehicle to identify and select among multiple potential coordination vehicles for a given maneuver. Multi-target maneuver coordination does not require modifications to the maneuver execution logic or to the underlying coordination protocol. Instead, it extends the decision-making process preceding coordination, enabling vehicles to exploit a broader set of feasible cooperative interactions. Results show that multi-target maneuver coordination significantly increases triggered and successfully executed coordinations while maintaining a low computational cost, as the proposed approach achieves these gains without requiring the analysis of a large number of potential target vehicles. These improvements preserve coordination success rates while enabling earlier maneuver initiation.**

## I. Introduction

Connected and Automated Vehicles (CAVs) can improve traffic safety and efficiency through cooperation enabled by Vehicle-to-Everything (V2X) communications. With maneuver coordination, vehicles can negotiate and cooperatively execute maneuvers that would otherwise be difficult, inefficient, or unsafe to perform. To this aim, vehicles can exchange information about their driving intentions, future trajectories, and maneuver constraints, in order to anticipate driving conflicts and actively collaborate to resolve them. Maneuver coordination is currently being standardized by ETSI TS 103 561 and SAE J3186. These standardization efforts define general coordination principles, message types, and a set of representative use cases such as cooperative lane change, cooperative lane merge, or intersection-related maneuvers. However, many critical design aspects that influence its effectiveness remain unspecified or underexplored.

Most existing research on maneuver coordination focuses on the definition of negotiation protocols [1][2], message exchange mechanisms [2][3], or the execution of specific cooperative maneuvers [4]. In these studies, the coordination process is typically framed as a bilateral interaction between an ego vehicle requesting coordination and a single target (or cooperating) vehicle, which is implicitly or explicitly pre-identified. In practice, this target vehicle is usually the one whose current position or trajectory directly prevents the execution of the desired maneuver of the ego vehicle [1]. If coordination with this specific vehicle is not feasible or fails, the maneuver is not performed and is either postponed or abandoned [3][4][5]. While this predefined selection of the target vehicle simplifies the coordination problem, it implicitly assumes that, at any given time, there is only one meaningful maneuver coordination opportunity, which limits the number of coordination opportunities.

Maneuver coordination inherently relies on anticipating the evolution of traffic over a future time horizon rather than reacting solely to the current instantaneous traffic state. By considering how surrounding vehicles are expected to move in the next few seconds, coordination decisions could be planned in advance, opening the possibility of alternative coordination opportunities that are not immediately available at the current time. From this perspective, it is possible that multiple surrounding vehicles may be capable of enabling the same maneuver, either through different coordination strategies or at different future time instants. For example, a vehicle intending to change lanes might be able to coordinate with the current vehicle immediately behind it in the target lane, or with the vehicle following it. Similarly, in merging or lane-drop scenarios, different vehicles may offer distinct opportunities to facilitate the maneuver depending on how traffic is expected to evolve.

In this context, this paper proposes a maneuver coordination approach based on multi-target selection, in which the ego vehicle does not assume a predefined cooperating vehicle, but instead identifies and selects a suitable coordinating vehicle among a set of potential target vehicles. Allowing a vehicle to consider and select among multiple target coordinating vehicles increases the probability that at least one feasible and beneficial coordination exists, and hence has the potential to augment the number of maneuver coordinations that can be successfully performed. Despite its potential, multi-target maneuver coordiations are not explicitly addressed in current standards. ETSI and SAE define the general concept of maneuver coordination and list representative use cases. However, ETSI and SAE standards do not specify how a vehicle requiring coordination should identify, evaluate, and select a target vehicle to coordinate with. Similarly, most existing studies implicitly assume a predefined target vehicle and do not analyze the impact of multiple potential target vehicles on the effectiveness of maneuver coordination [3][4][5]. In this paper, we propose and investigate the potential of multi-target maneuver coordinations. We study how the ability to select among several potential target vehicles for coordination influences the number of successful maneuver coordinations. We quantify the impact of multi-target maneuver coordinations, and demonstrate that the proposed approach allows vehicles to make more informed coordination decisions and exploit a broader set of feasible cooperative interactions. The results demonstrate that multi-target maneuver coordinations can increase the number of successfully executed maneuver coordinations without degrading the success rate.

## II. Maneuver coordination design

We follow ETSI and SAE principles to design maneuver coordinations for realistic traffic conditions.

### A. State machine, vehicle roles and messages exchange

We implement maneuver coordinations as a distributed protocol based on a state machine (Fig. 1), where vehicles exchange coordination-related messages and assume specific roles during the process [5]. When not actively engaged in a coordination, vehicles operate in an *Intent-Sharing* state (referred to as the "pre-awareness state" in SAE terminology). In this state, they periodically broadcast *Intent* messages describing their planned trajectory, which represents the path they intend to follow over the next few seconds[1]. These messages allow surrounding vehicles to continuously assess the traffic context and anticipate potential conflicts or cooperation opportunities. To limit channel load while preserving awareness, *Intent messages* are generated following [6] with a maximum interval of 1 s.

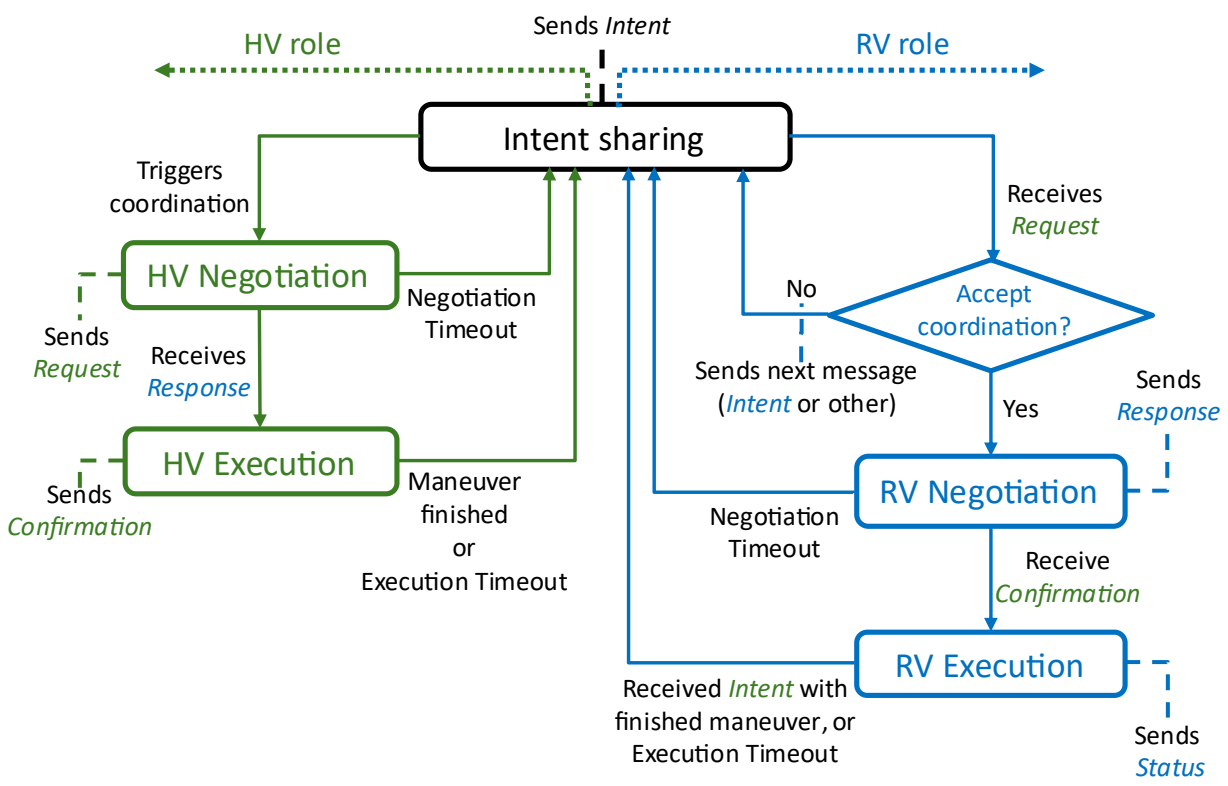

Fig. 1. State machine model for maneuver coordination.

When a vehicle determines that coordination is required, it assumes the role of Host Vehicle (HV) and initiates the coordination process. The target vehicle with which the HV attempts to coordinate is referred to as the Remote Vehicle (RV). The coordination itself is structured into two phases, negotiation and execution. During the Negotiation phase, the HV transitions to an *HV Negotiation* state and repeatedly broadcasts *Request* messages every 100 ms. These messages describe the maneuver to be coordinated and include relevant timing information. The HV stays in the *HV Negotiation* state until either the RV responds with a *Response* message accepting the request, a different message implicitly rejecting the coordination, or the *Negotiation Timeout* expires. The *Negotiation Timeout* is predefined at a fixed time measured after the Coordination Triggering (CT) time. On the other side, upon receiving a *Request*, a vehicle decides whether it can participate as the RV. If it declines (e.g., because it is already involved in another coordination), it remains in its current state without changing its planned trajectory and promptly broadcasts its next message, implicitly signaling rejection of the coordination request. If it accepts, it transitions to the *RV Negotiation* state and transmits *Response* messages every 100 ms until the negotiation is finalized by the reception of a *Confirmation* message by HV, or until the *Negotiation Timeout* expires.

A successful negotiation leads both vehicles into the Execution phase. Upon the reception of the *Response* message, the HV enters the *HV Execution* state and starts broadcasting *Confirmation* messages, again every 100 ms, to confirm that the maneuver execution is ongoing. Upon the reception of the *Confirmation* message, the RV enters the *RV Execution* state and starts broadcasting maneuver coordination *Status* messages, also every 100 ms, to acknowledge the ongoing execution on the RV side as well. During the execution phase, both vehicles perform the agreed actions until the maneuver is completed or an *Execution Timeout* is reached. This timeout is computed from the estimated Coordination Intended Finish (CIF) time plus a predefined margin to account for execution deviations. Once execution ends, both vehicles return to the *Intent-Sharing* state.

All coordination-related messages (*Request, Response, Confirmation, Status*) include the HV and RV identifiers, the maneuver ID, and the CT and CIF times, ensuring proper synchronization and prevent ambiguities when multiple coordinations coexist.

### B. Maneuver coordination triggering and execution

Before initiating a maneuver coordination, a vehicle must determine whether the coordination is both beneficial and feasible. This evaluation is performed continuously by each vehicle based on its perception of the traffic environment and its ability to anticipate how traffic will evolve over the next few seconds. A vehicle can anticipate the evolution of traffic thanks to *Intent-Sharing*. By combining its own planned trajectory with the *Intent-Sharing* received from surrounding vehicles, a vehicle estimates the future trajectories of relevant neighbors over a short horizon (typically few seconds). Based on the predicted traffic state, the vehicle evaluates whether a specific maneuver would be desirable but cannot be safely or efficiently executed without cooperation. If so, it assesses whether the maneuver could become feasible with the cooperation of another vehicle. Only if both benefit and feasibility conditions are satisfied, the vehicle triggers the maneuver coordination process.

Once coordination is triggered and successfully negotiated, vehicles execute the maneuver. The specific actions of HV and RV are maneuver-dependent but must be specified in advance to ensure safety and predictability. In general terms, the HV executes the desired maneuver while the RV performs a complementary action that facilitates it, e.g. adjusting speed or position. To illustrate this process, we consider a coordinated lane change maneuver. In this case, the HV intends to change lanes but is blocked by traffic in the target lane. The RV is a vehicle that can make the lane change feasible by creating a gap through a controlled deceleration. After negotiation, the RV decelerates for a limited time while the HV monitors whether the lane change remains safe and beneficial. The HV executes the maneuver once a sufficient gap is created and safety and efficiency criteria are met. The same coordination logic applies to other scenarios, such as merging or intersections, with different execution patterns.

[1] Other messages such as CAMs (Cooperative Awareness Messages) and CPMs (Collective Perception Messages) may also be transmitted.

## III. MULTI-TARGET MANEUVER COORDINATION

Maneuver coordination is inherently based on anticipating the future evolution of traffic. As a result, coordination opportunities should not be necessarily limited to a single surrounding vehicle at a given time. Instead, multiple vehicles may potentially enable the same maneuver through different coordination strategies or at different future time instants. This observation motivates extending maneuver coordination beyond the conventional predefined target approach. In most existing maneuver coordination studies, once a vehicle determines that coordination is required, a single coordination vehicle is implicitly assumed [1]. This vehicle is typically the one whose current position or planned trajectory directly prevents the execution of the desired maneuver. If coordination with this specific vehicle is not feasible, maneuver coordination is not triggered, and the maneuver is postponed or abandoned. While this approach is simple and intuitive, it restricts coordination to a single opportunity and overlooks alternative coordination possibilities that may arise when considering traffic evolution over time. This limitation is clearly illustrated in the coordinated lane change maneuver represented in Fig. 2. In this maneuver, an ego vehicle $V_{ego}$ determines, based on its lane change incentive logic, that it wants to move to an adjacent lane. However, the desired trajectory is in conflict with the current planned trajectory of the vehicle immediately behind in the target lane, denoted as $V_{bn1}$. To resolve this conflict, $V_{ego}$ attempts to coordinate with $V_{bn1}$ to create the required gap for a safe lane change through controlled deceleration. If this deceleration is not feasible, e.g., because $V_{bn1}$ is traveling much faster, coordination is not triggered and $V_{ego}$ must wait until traffic conditions change.

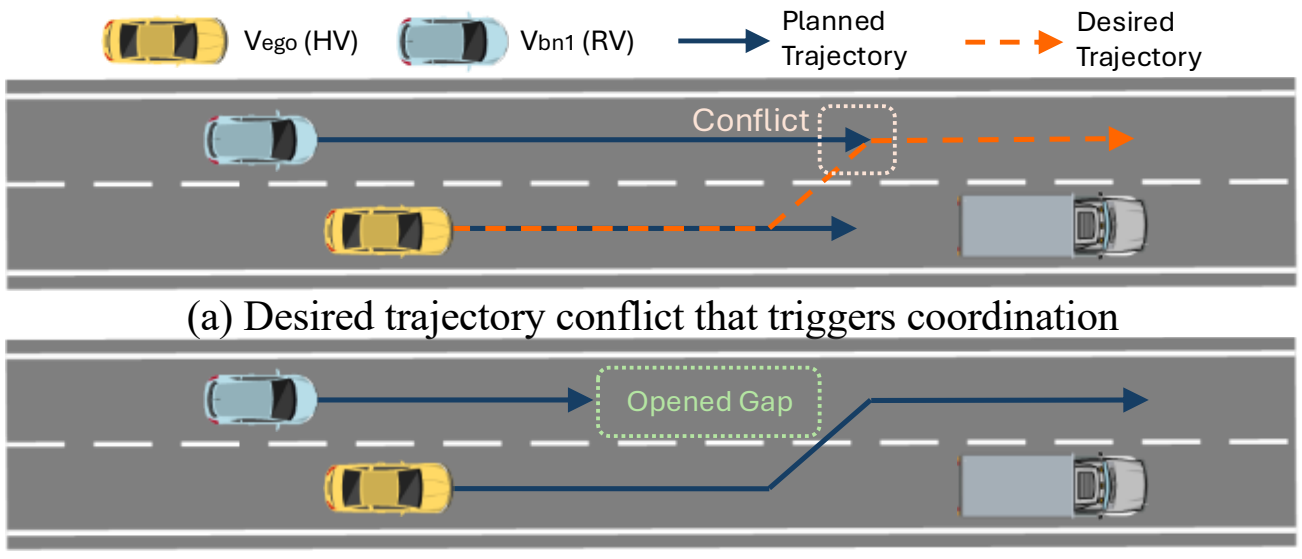

(a) Desired trajectory conflict that triggers coordination

(b) Coordinated new planned trajectories

Fig. 2. Maneuver coordination using a fixed predefined target.

However, when *Intent-Sharing* is available, $V_{bn1}$ is not necessarily the only vehicle that could facilitate the maneuver. Other vehicles in the target lane, such as the vehicle traveling behind $V_{bn1}$ ($V_{bn2}$), or even further vehicles ($V_{bn3}$, $V_{bn4}$, etc.), may also represent valid coordination partners for the ego vehicle. By coordinating with one of these vehicles, a gap can be created with greater anticipation and exploited by $V_{ego}$ once it reaches the appropriate longitudinal position. Importantly, the coordinated maneuver itself remains unchanged: the objective is still to create a sufficient gap in the target lane through a controlled deceleration by a vehicle in that lane. What differs is the selected target vehicle and the timing of the maneuver execution. Fig. 3 illustrates this alternative coordination strategy. In Fig. 3a, $V_{ego}$ identifies a conflict between its desired trajectory and the planned trajectory of $V_{bn1}$ that cannot be solved safely with a maneuver coordination with $V_{bn1}$, e.g. because it would require a strong deceleration of $V_{bn1}$. Alternatively, $V_{ego}$ could coordinate the lane change maneuver with $V_{bn2}$ Fig. 3b if the conflict between $V_{ego}$'s desired trajectory and $V_{bn2}$'s planned trajectory is solved. Fig. 3c shows the planned trajectories of $V_{ego}$ and $V_{bn2}$ after negotiation to create the required gap for the lane change. Motivated by the advantages illustrated in this example, we propose multi-target maneuver coordination, enabling an ego vehicle to select among multiple potential coordinating vehicles instead of a single predefined one.

The multi-target maneuver coordination maintains the coordination protocol described in Section II. It follows the same procedure regardless of which target vehicle is selected. The difference lies exclusively in the decision-making stage, where multiple target vehicles are evaluated and one is selected for coordination. The criteria used to select the coordinating vehicle depends on the maneuver type and may involve trade-offs between feasibility, timeliness, and traffic impact. In this work, we consider a coordinated lane change maneuver and adopt a simple and intuitive selection policy. The ego vehicle first assesses the feasibility of coordinating with $V_{bn1}$, the vehicle immediately behind in the target lane, as this typically results in the earliest maneuver execution. If coordination with $V_{bn1}$ is not feasible, the ego vehicle sequentially evaluates the following vehicles immediately behind in the target lane, i.e. $V_{bn2}$, followed by $V_{bn3}$ and $V_{bn4}$ in Fig. 3, if necessary, up to a maximum number of potential target vehicles. This sequential selection is compatible with existing maneuver coordination protocols and standards while enabling additional coordination opportunities.

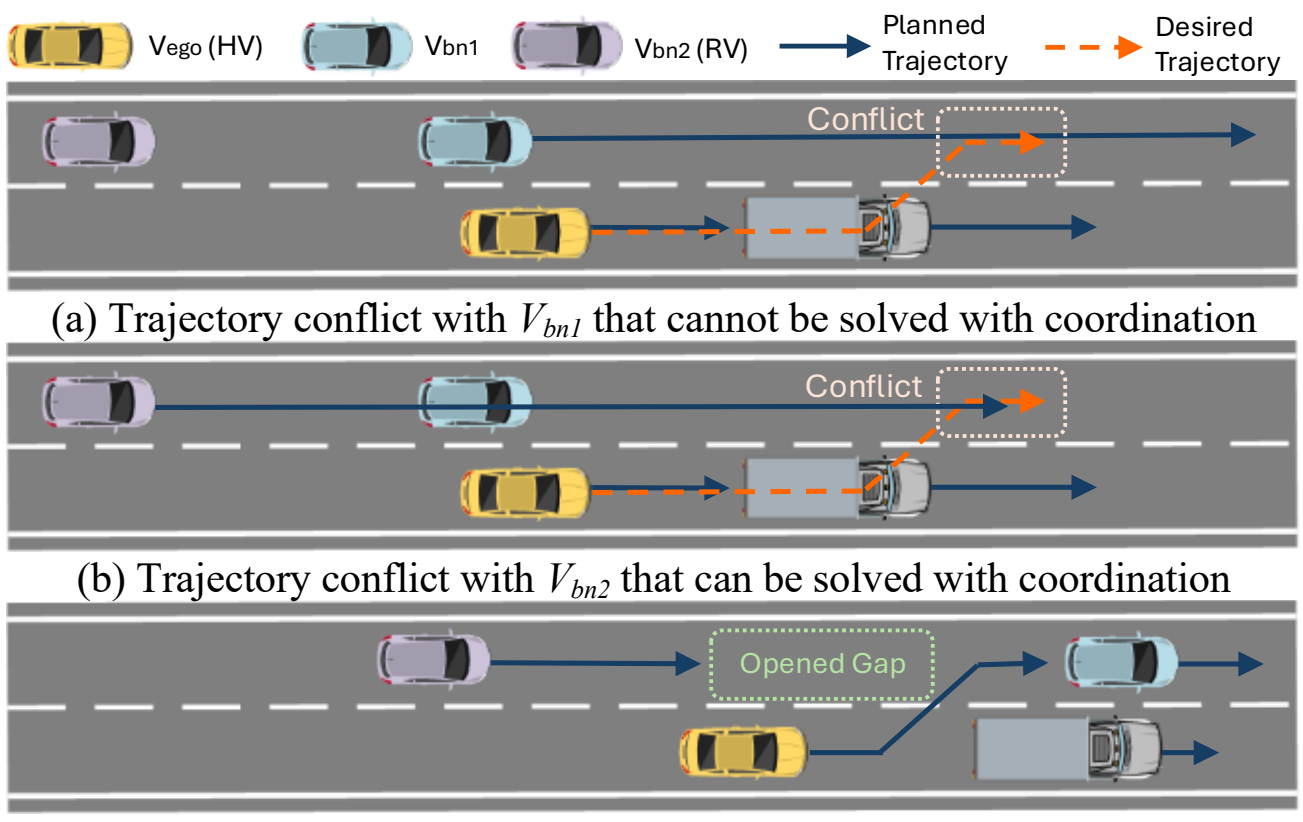

(a) Trajectory conflict with $V_{bn1}$ that cannot be solved with coordination

(b) Trajectory conflict with $V_{bn2}$ that can be solved with coordination

(c) Coordinated lane change with $V_{bn2}$

Fig. 3. Multi-target coordinated lane change.

## IV. SIMULATION FRAMEWORK

We evaluate the impact of the proposed multi-target maneuver coordination using our integrated traffic and V2X simulation framework [5][7]. V2X communication is simulated using the ns-3 network simulator. Vehicle mobility is modeled using a modified version of the VANET Highway Mobility module for ns-3 [8]. Simulations are conducted on a 5-km highway with six lanes (three per direction) using periodic boundary conditions, where exiting vehicles reappear at the opposite end with the same lane and speed. Performance is evaluated under five traffic densities. Traffic includes 80% passenger cars and 20% trucks, with desired speeds uniformly distributed around 120 km/h for cars and 80 km/h for trucks (±20%). Trucks are restricted to the middle and right lanes.

Each scenario is evaluated through 20 independent 200-s simulations to ensure statistical significance.

We focus on coordinated lane changes and enforce a minimum 400 m separation between active coordinations to avoid interference. Consequently, a vehicle cannot trigger coordination if another is active within this range. The Negotiation Timeout is fixed at 500 ms after Coordination Triggering (CT) time. Trajectories are computed over a 5 s prediction horizon, defining the maximum Coordination Intended Finish (CIF) time. An additional 3 s margin determines the Execution Timeout, yielding a maximum coordination duration of 8 s from triggering to completion. During execution, the RV decelerates at -2 m/s² for 1 s and then maintains constant speed. The HV follows its planned trajectory and changes the lane once a sufficient gap is available and safety and efficiency conditions are met.

## V. Results and Discussion

Fig. 4 shows the number of triggered lane change maneuver coordinations per vehicle per hour for different coordination configurations. The baseline configuration corresponds to the conventional predefined target approach, in which coordination is only attempted with the vehicle immediately behind in the target lane (i.e. RV is always $V_{bn1}$). The remaining curves correspond to the configurations where the potential coordination targets are: $V_{bn1}$ and $V_{bn2}$; $V_{bn1}$, $V_{bn2}$ and $V_{bn3}$; $V_{bn1}$, $V_{bn2}$, $V_{bn3}$ and $V_{bn4}$. The results in Fig. 4 show a clear and relevant increase in the number of triggered coordinations when comparing the single-target coordination with the multi-target coordination including $V_{bn2}$. This indicates that a substantial fraction of coordination opportunities that cannot be realized with $V_{bn1}$ were actually feasible when a second potential target vehicle was considered. In contrast, enabling coordination with additional vehicles beyond $V_{bn2}$ does not provide significant gains for the current scenario. The increase observed when including $V_{bn3}$ is marginal, and no further increase is observed when $V_{bn4}$ is also enabled. This is the case because most maneuver coordinations finally only involve $V_{bn1}$ and $V_{bn2}$. This is visible in Fig. 5, which plots the distribution of triggered maneuver coordinations across possible target vehicles ($V_{bn1}$ to $V_{bn4}$). The figure reveals that 70-80% of the triggered coordinations involve $V_{bn1}$, while 20-30% are performed with $V_{bn2}$. Coordinations involving $V_{bn3}$ account for only about 0.5% of the total, and no coordinations with $V_{bn4}$ are observed. The negligible contribution of $V_{bn3}$ and $V_{bn4}$ can be explained by the limited difference in average speed between lanes in the evaluated scenarios. In most cases, a gap created by a vehicle located far behind in the target lane cannot propagate forward to the ego vehicle's position given the duration of the considered planned trajectories (5 s). As a result, coordination with distant vehicles rarely meets the feasibility conditions required to trigger coordination. Multi-target maneuver coordination increases triggered coordinations without requiring the ego vehicle to analyze many potential targets.

Fig. 6 compares the percentage of successful maneuver coordinations achieved with the conventional single-target approach and the proposed multi-target scheme, which considers four potential target vehicles ($V_{bn1}$-$V_{bn4}$). Results show that multi-target coordination maintains a success rate comparable to the single-target approach, indicating that enabling coordination with more distant vehicles does not significantly increase unsuccessful executions. This holds despite the higher sensitivity to trajectory errors caused by traffic variations when coordinating with distant target vehicles. Small deviations in intermediate vehicle behavior or traffic dynamics more strongly affect maneuvers relying on gaps created further upstream in the target lane. Nevertheless, multi-target coordination achieves success rates comparable to the conventional single-target approach.

The combined effect of the results shown in Fig. 4 and Fig. 6 is reflected in Fig. 7, which reports the number of successful maneuver coordinations per vehicle per hour. Fig. 7 shows the increase in triggered coordinations achieved with multi-target coordinations translates almost proportionally into an increase in successful coordinations. This confirms that the additional coordination opportunities enabled by multiple target vehicles do not reduce execution reliability. Overall, these results show that multi-target coordination increases both triggered and successful maneuvers.

Fig. 8 reports the coordination anticipation time, defined as the average time difference between the Coordination Intended Finish (CIF) time and the Coordination Triggering (CT) time. This metric reflects how far in advance maneuver coordinations are initiated on average. The results compare the conventional single target approach with multi-target coordination approach. For the multi-target coordination approach we differentiate the coordinations with $V_{bn1}$, $V_{bn2}$ and any $V_{bn}$. The results in Fig. 8 show a clear difference in the anticipation level of maneuver coordinations depending on the selected coordination vehicle. Coordinations performed with $V_{bn1}$ exhibit the smallest coordination anticipation time, reflecting the fact that these coordinations are typically triggered close to the moment when the maneuver can be executed. In contrast, coordinations involving $V_{bn2}$, $V_{bn3}$, or $V_{bn4}$ are initiated significantly earlier, resulting in a larger coordination anticipation time. This behavior is a direct consequence of the coordination strategy considered. When coordinating with vehicles farther behind in the target lane, the creation of a suitable gap is initiated earlier so that the gap can propagate forward. As a result, these coordinations inherently rely on greater anticipation. This result highlights that, beyond increasing the number of coordinated maneuvers, multi-target coordination also enables coordinations to be triggered earlier, reinforcing its potential benefits due to better anticipation.

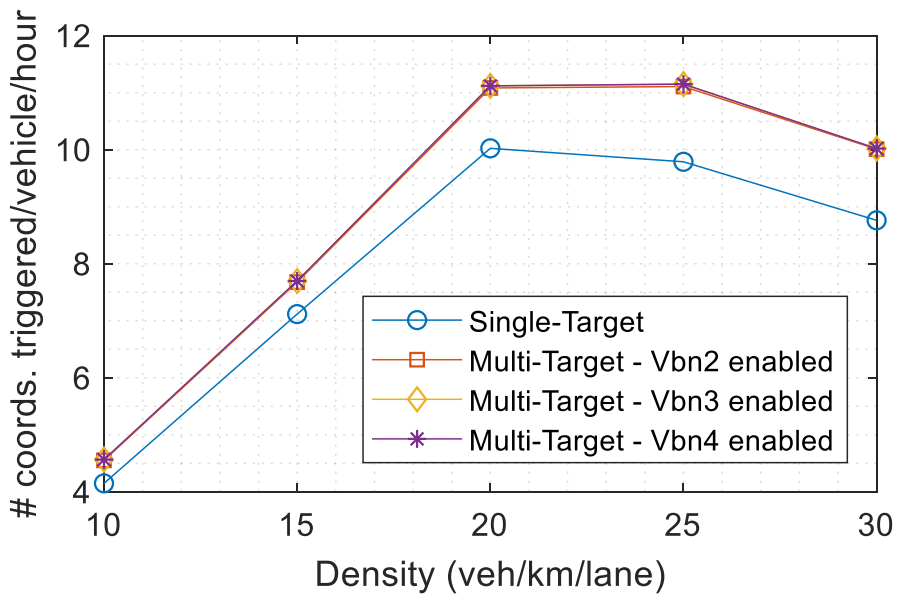

Fig. 4. Number of triggered lane change coordinations per vehicle per hour for single and multi-target maneuver coordinations.

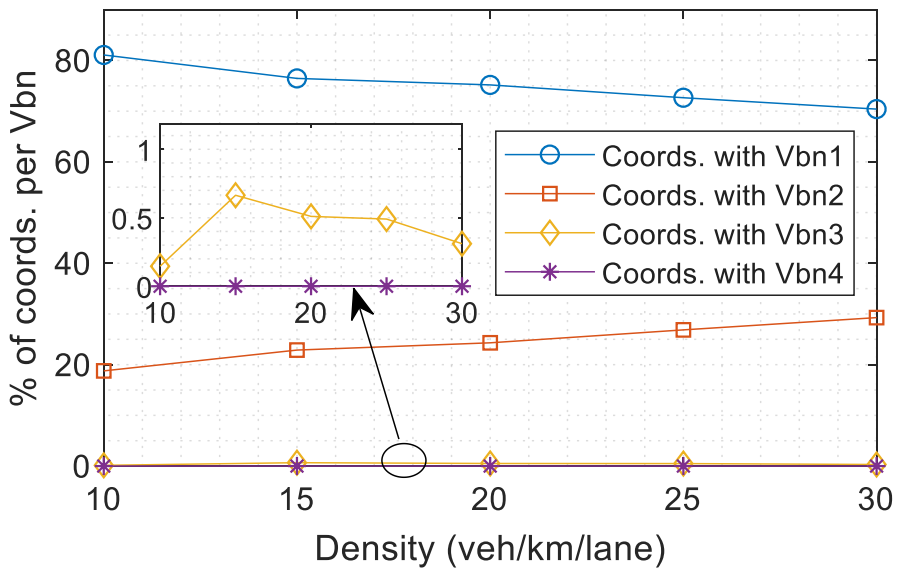

Fig. 5. Distribution of triggered maneuver coordinations across target coordination vehicles for multi-target coordination.

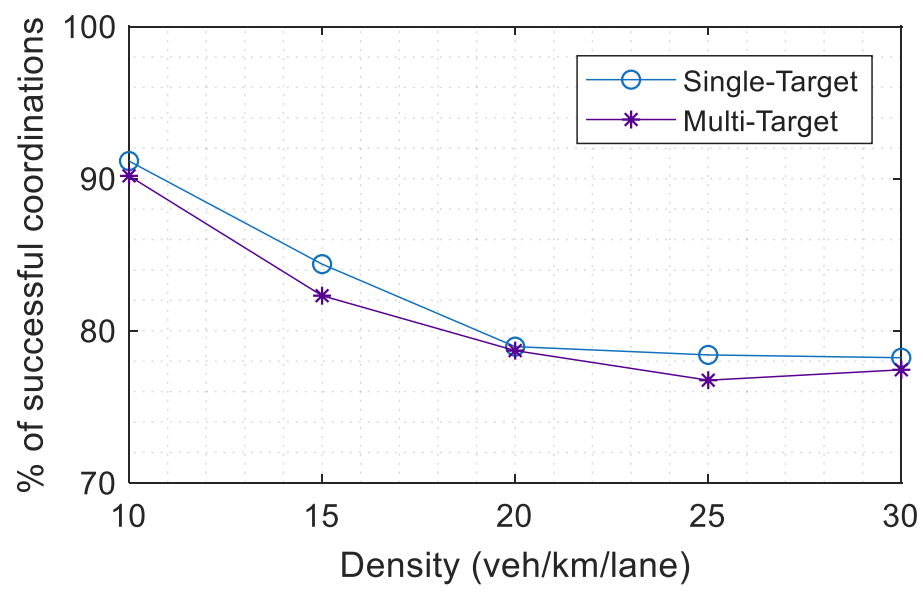

Fig. 6. Percentage of successful maneuver coordinations for single and multi-target maneuver coordinations.

We evaluate the performance of the multi-target coordination proposal in a more demanding traffic scenario where a stationary obstacle is placed in the rightmost lane of each driving direction. The obstacle forces right-lane vehicles to change lanes, increasing lane-change demand and potentially favoring coordinations requiring greater anticipation. Fig. 9 shows the distribution of triggered maneuver coordinations across coordination targets in this new scenario, following the same representation used in Fig. 5. The results indicate a slight shift in the distribution of coordination targets when compared to the scenario without obstacles. In particular, the proportion of coordinations performed with $V_{bn3}$ increases to approximately 1%, and a small number of coordinations with $V_{bn4}$ are observed. These results suggest that the presence of an obstacle can indeed promote coordination opportunities that involve more distant vehicles in the target lane. Despite this effect, the overall trends remain consistent with those observed in the obstacle-free scenario. Coordinations with $V_{bn1}$ and $V_{bn2}$ still dominate with no significant differences in the remaining metrics.

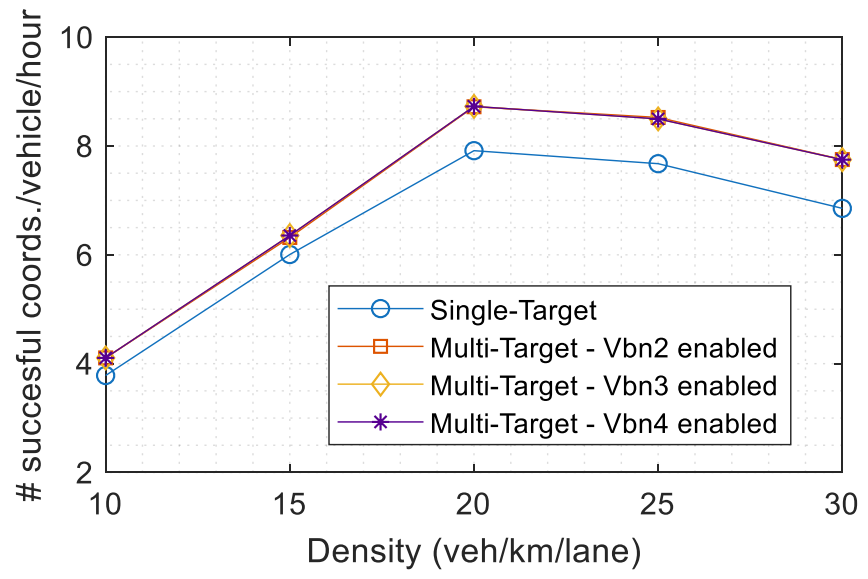

Fig. 7. Successful maneuver coordinations per vehicle per hour for single and multi-target maneuver coordination.

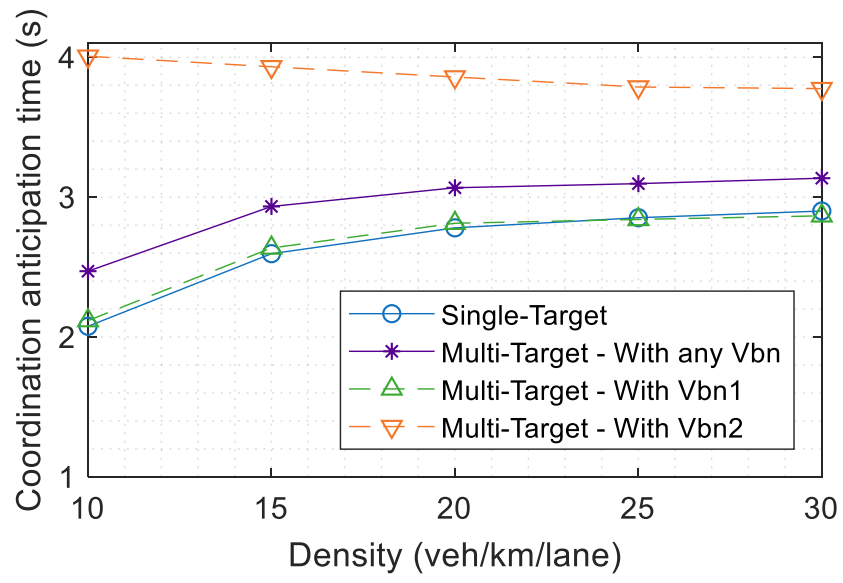

Fig. 8. Coordination anticipation time.

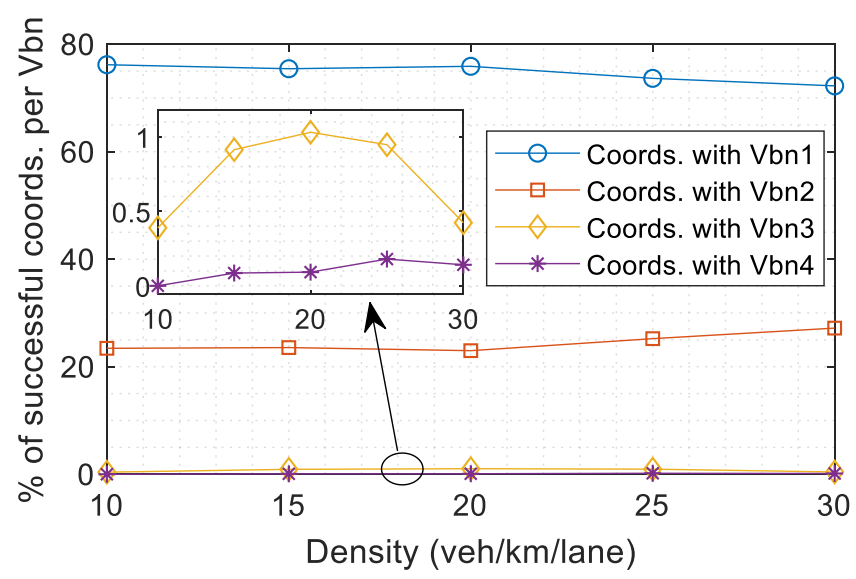

Fig. 9. Distribution of triggered maneuver coordinations across target vehicles for multi-target coordination in the obstacle scenario.

## VI. Conclusions

This paper proposed and analyzed the impact of multi-target maneuver coordinations, where a vehicle can select among several potential target vehicles. We demonstrated that considering multiple target candidates increases the number of triggered and successfully executed maneuver coordinations. These gains are achieved without compromising coordination reliability, and multi-target maneuver coordinations achieve similar maneuver coordination success rates as the conventional single and predefined target maneuver coordinations. Results have also shown that most of the observed gains are achieved only considering a few target vehicles, which reduces implementation complexity. In addition, multi-target maneuver coordination enables maneuvers to be triggered with greater anticipation, providing additional temporal flexibility to cope with traffic dynamics and uncertainty. The target selection criterion adopted in this work is a simple and intuitive first approach that illustrates the potential gains of multi-target maneuver coordination. Although this policy already provides significant benefits, more advanced strategies could be explored in future work, including adaptations to specific traffic conditions or maneuver types. Overall, multi-target maneuver coordination emerges as a simple yet effective enhancement that amplifies coordination benefits without modifying the underlying protocol or maneuver implementation. This opens the door to future research on cooperator selection strategies and more complex multi-maneuver scenarios.

## Acknowledgements

Work partially funded by MICIU/AEI/10.13039/501100011033 and "ERFD/EU" (PID2023-150308OB-I00), and Generalitat Valenciana (CIAICO/2024/167).